\documentstyle[twocolumn,prl,aps,psfig]{revtex}
\tightenlines
\begin{document}
\draft
\title{Spin-Orbit induced semiconductor spin guides}
\author{Manuel Val\'{\i}n-Rodr\'{\i}guez, Antonio Puente
and Lloren\c{c} Serra}
\address{Departament de F\'{\i}sica, Universitat de les Illes Balears,
E-07071 Palma de Mallorca, Spain}
\date{November 7, 2002}
\maketitle
\begin{abstract}
The tunability of the Rashba spin-orbit coupling allows to build
semiconductor heterostructures with space modulated coupling intensities.
We show that a wire-shaped spin-orbit modulation in a quantum well  
can support propagating electronic states inside the wire
only for a certain spin orientation and, therefore, it acts as an 
effective spin transmission guide for this particular spin orientation. 
\end{abstract}

Currently, one of the most challenging issues in condensed matter  
physics is the injection and control of the electronic spin in 
semiconductor heterostructures. Its principal interest comes from the 
applications point of view, since the electron spin in semiconductor 
nanostructures has revealed as a promising candidate to implement 
quantum bits, a necessary ingredient of quantum computation \cite{loss98}.
The feasibility of spin-based electronic 
devices \cite{wolf01,datt90,koga02,gove02} 
also relies on the ability to manipulate the spin carriers.

In recent years, spin polarization has been induced in semiconductors
using optical \cite{kikka98} and electrical methods \cite{fied99,zhu01}.
In the case of electrical injection, the technique relies on external
`spin-alligner' elements, such as magnetic semiconductors 
or ferromagnetic metals, as a previous step that polarizes the current 
which is injected into the semiconductor. 
In this work we propose an alternative mechanism that internally selects 
the propagating spins in the semiconductor by means of a space
modulation in the Rashba spin-orbit (SO) coupling for heterostructures with 
inversion asymmetry. 

We consider a two-dimensional electron gas (2DEG) confined to
a III-V quantum well whose
inversion asymmetry produces an electric field in the 
perpendicular ($z$) direction. As a consequence of the relativistic 
corrections this electric field acts on the 2DEG carriers as an
effective SO coupling field known as the 
Rashba term \cite{ras60,kna96}. 
The strength of the Rashba coupling depends on the heterostructure's 
vertical electric field and has been 
shown to be experimentally controllable with a tunable gate 
voltage \cite{nitt97}.
We take advantage of this tunability
to define a heterostructure with a space modulated SO coupling intensity,
depicted schematically in the upper part of Fig.\ 1. 
In practice, the SO coupling variation 
would correspond to a modulated electric field in the vertical 
direction. The well has a constant SO coupling strength 
$\lambda_R^{(e)}$ except within a narrow region (guide) of width $a$ 
where it takes the `internal' value $\lambda_R^{(i)}$. 

We model the conduction electrons of this 
heterostructure using the effective mass Hamiltonian  
\begin{eqnarray}
\label{eq1}
{\cal H} =\frac{p_x^2+p_y^2}{2m^*} + \frac{\lambda_R(x)}{\hbar} 
\left(\, p_y\sigma_x-p_x\sigma_y\,\right) \; ,
\end{eqnarray}
where the modulated SO coupling reads
\begin{eqnarray}  
\lambda_R(x)=\left\{\begin{array}{c}
	\lambda_R^{(i)} \;\;{\rm if}\;\;  |x| \leq\frac{a}{2} \\[0.1cm]
	\lambda_R^{(e)} \;\;{\rm if}\;\;  |x| >\frac{a}{2}
\end{array}\right.\; .
\end{eqnarray}  
Equation (\ref{eq1}) also contains the conduction band effective mass $m^*$ 
and the Pauli matrices $\sigma_x$ and $\sigma_y$ corresponding to
the in-plane electron spin.

Since the Hamiltonian is translationally invariant in
$y$ coordinate its eigenstates have well-defined momentum in that direction. 
In this representation the Hamiltonian is separable in space coordinates 
and the eigenstates are composed of a propagating longitudinal plane-wave, 
having $y$ momentum ${\hbar}k$, and a spinorial transverse profile
\begin{equation}
 \left(\begin{array}{c}
	\psi_{nk\uparrow}({\bf r}) \\
	\psi_{nk\downarrow}({\bf r}) 
	\end{array}
	\right) = e^{iky}
 \left(\begin{array}{c}
	\phi_{nk\uparrow}(x) \\
	\phi_{nk\downarrow}(x)
	\end{array}\right)\; ,
\end{equation}
where the spinorial part is given in the usual $\sigma_z$ basis.

From the experimental data reported by
Nitta et al. \cite{nitt97} for an 
In$_{0.53}$Ga$_{0.47}$As/In$_{0.52}$Al$_{0.48}$As heterostructure,
we extract the following parameter values, to be used below
in the numerical applications:
$m^*=0.05\, m_e$, where $m_e$ is the free electron mass;
$\lambda_R^{(i)}=0.5\times 10^{-9}$ eVcm; 
$\lambda_R^{(e)}=1.0\times 10^{-9}$ eVcm; relative dielectric constant
of InGaAs $\epsilon=13.9$; 
$y$-coordinate wavevector $k=3.5\times 10^6$ cm$^{-1}$, 
near the Fermi wavevector of a 2DEG with density
$n_s\simeq 2\times 10^{12}$ cm$^{-2}$. 

Before presenting numerical results with the above Hamiltonian
and in order to clarify our purpose we consider the 
following simplification on Eq.\ (\ref{eq1}): 
let us assume that the 
guide can have localized states in the $x$ coordinate
having a characteristic transverse wavevector 
$k_x\sim {\pi}/{a}\ll k$. 
Therefore, in a qualitative analysis, we can ignore 
the term containing $p_x$.
The eigenstates of the resulting approximated Hamiltonian have well 
defined spin in the $x$ direction and feel an effective spatial 
potential that depends on the $\sigma_x$ eigenvalue, i.e., it changes 
sign from $+x$ to $-x$ spin orientation.
Figure 1 depicts a schematic representation of the up and down 
potentials.
It has to be noted that the depth of the effective potential well 
for $+x$ and the height of the barrier for $-x$ spin orientation 
depend both on the spin-orbit modulation step 
$\Delta\lambda_R=\lambda_R^{(e)}-\lambda_R^{(i)}$ and the 
longitudinal $y$ wavevector of the state $k$. As $k$ increases the 
attractive or repulsive character will also increase linearly.

The above analysis indicates that transverse 
space-confined modes can exist only for one of the two spin orientations
thus showing the feasibility of selective spin guiding in this kind of 
heterostructures. We proof below that this conclusion remains valid
when the $p_x\sigma_y$ term in Eq.\ (\ref{eq1}) is taken into account. 
The $y$-translational invariance of the eigenstates 
makes the initial two-dimensional Hamiltonian in coordinate space reduce 
to an effective one-dimensional one, corresponding to the 
transverse modes
\begin{equation}
 {\cal H}_{\em tr} =  
-\frac{\hbar^2}{2m^*} \left(\frac{\partial^2}{\partial x^2}-k^2\right) 
+\lambda_R(x)\left(i\frac{\partial}{\partial x}\sigma_y 
+k\sigma_x \right)\; . 		
\end{equation}

We shall numerically obtain the relevant
transverse eigenmodes of the Hamiltonian ${\cal H}_{\em tr}$ from the 
resolution of the time-dependent Schr\"odinger equation 
uniformly discretized in the $x$ coordinate and in time.
The procedure is as follows: we evolve in time an initial spinorial 
wavepacket and, taking advantage of the harmonic time evolution for 
eigenmodes, we extract the transverse eigenspinors and eigenenergies 
using Fourier analysis of the time signals. 
For a certain value of $y$-wavevector $k$ the time evolution of any 
wavepacket is decomposed as:
\begin{equation}
 \left(\begin{array}{c}
	\varphi_{k\uparrow}(x,t) \\
	\varphi_{k\downarrow}(x,t) 
	\end{array}
	\right) = \sum_{n} A_{nk}
 \left(\begin{array}{c}
	\phi_{nk\uparrow}(x) \\
	\phi_{nk\downarrow}(x)
	\end{array}\right) e^{-i\omega_{nk}t}\; .
\end{equation}

Besides the eigenenergies $\omega_{nk}$ of the different transverse
modes $n$, Fourier transform at each grid point yields the local 
value of the eigenspinors. This method is appropriate to the present 
problem since an initial wavepacket inside the guide (see Fig.\ 1)
will excite the confined transverse modes, if they exist, 
or it will quickly spread to the bulk surrounding the guide.
We use Gaussian-shaped wavepackets in coordinate space with spin oriented 
in $x$ direction, i.e.,
\begin{equation}
\label{eq6}
{\bf \varphi}(x,t=0)\equiv
\frac{1}{2\sigma\sqrt{\pi}}\,e^{\frac{-x^2}{2\sigma^2}}
\chi_{\pm}^{(x)}\; ,
\end{equation}
with the $\sigma_x$ eigenspinors
\begin{equation}
\chi_\pm^{(x)} =
\left(\begin{array}{r}
		1 \\
	     \pm 1
      \end{array}\right)\; .
\end{equation}
The spatial spread of the wavepacket is chosen close to the width of the
spin-orbit guide because we want a maximum overlap with the confined 
transverse modes. Gaussian wavepackets are appropiate to excite
confined modes having even parity in the $x$ coordinate while for
odd parity states it is convenient to use an antisymmetric wavepacket.
This can be done by multiplying Eq.\ (\ref{eq6}) by $x$.

Physically, the input wavepacket represents an electron having 
well-defined $p_y$ and spin oriented along $x$, injected 
from a lead with transversal dimensions matching those of the 
guide. 
Figure 2 shows the time evolution of the density for initial 
wavepackets with two different initial spin orientations: 
$+x$ (up) and $-x$ (down), using a guide width $a=60$ nm.
It can be seen that for down initialization the wavepacket spreads over 
the transversal dimension in $t\sim 1$ ps,  with a complete depletion 
of density inside the spin guide. For spin up a part of the probability
density remains confined to the guide as time evolves. This means 
that our initial up state was composed of confined and travelling transversal 
modes, while the down state is only composed of the latter ones. 
For a general orientation of the initial spin the confined fraction of 
the density depends on $\langle\sigma_x\rangle$, vanishing if 
the initial spin points in the $-x$ direction and reaching 
a maximum for the $+x$ one.

To identify and extract the confined modes present in the time evolution 
simulation we use
the mentioned Fourier analysis technique, focussing on a frequency 
region of the order $\Delta\lambda_R k$.
This analysis reveals that for our parameter set there is only one confined
mode with spin polarization mainly oriented in $+x$ direction, as 
was expected from the above discussion.
To observe a second confined mode (antisymmetric), maintaining the width 
of 60 nm, the longitudinal wavevector should be increased up to 
$k\sim 6\times 10^6$ cm$^{-1}$. 
Actually, for a fixed $\Delta\lambda_R$ the number of confined states 
depends only on the product $ka$. 
The probability and spin densities of the confined mode, related to the 
wavefunction by
\begin{eqnarray}
\langle\rho\rangle(x) &=&  \mid\!\phi_{n k\uparrow}(x)\!\mid^2
+\mid\!\phi_{nk\downarrow}(x)\!\mid^2 
\nonumber\\
\langle\sigma_z\rangle(x) &=& \mid\!\phi_{n k\uparrow}(x)\!\mid^2
-\mid\!\phi_{nk\downarrow}(x)\!\mid^2
\nonumber\\
\langle\sigma_x\rangle(x) &=& 2 {\rm Re}\left\{
\phi_{n k\uparrow}(x) \phi_{n k\downarrow}^*(x) \right\}\; ,
\end{eqnarray}
are shown in Fig.\ 3.

Probability and spin densities clearly show the confining character
of the guided mode, the main spin polarization
being concentrated in $+x$ direction. The mode has zero 
$\langle\sigma_y\rangle(x)$ and a small $\langle\sigma_z\rangle(x)$. 
This confined mode is not an eigenstate of $\sigma_x$, as the transversal 
Hamiltonian ${\cal H}_{\em tr}$
does not commute with this operator but, nevertheless, it has an 
expectation value $\langle\sigma_x\rangle= 0.99$, very close to
the value $1$ for eigenstates.

It is worth to mention that since time-reversal symmetry is conserved 
by the Hamiltonian of Eq.\ (\ref{eq1}), Kramers degeneracy must hold.
Therefore, a counterpart to the confined mode discussed above exists. 
This complementary state is also confined to the guide but its 
spin polarization and longitudinal momentum ($p_y$) are inverted,
i.e., it is mainly oriented in $-x$ direction and has $k'=-k$.

The ability of the guide to confine additional modes is enhanced 
by increasing the guide-width $a$. Figure 4 shows the evolution 
with $a$, maintaining $k$ to the same value as before. 
Confined modes should lie between the energy edges of the guide,
marked in Fig.\ 4 by dotted lines.
We note that the second and third modes appear for guide widths of 
$a\approx 80$ and $a\approx 150$ nm, respectively, with probability 
densities having an additional node for each successive mode.
As the first $n=0$ state these higher modes have spin almost 
completely oriented in $+x$ direction.

The confining energies of the 
guide are of order $\Delta\lambda_{R}k$, which takes a value 
of $\sim 2$ meV for our parameter set. Consequently, the practical operation 
of the proposed guide is limited to low temperatures, since the coupling
with phonons in the meV range would reduce the efficiency of 
the guiding effect. 
In this sense, it has to be pointed out that the assignment of a lower 
spin-orbit constant to the guide region
is not arbitrary, because, as reported by Nazarov {\em et al.} \cite{Kha01}, 
spin-orbit coupling induces an admixture of
pure up and down spin orientations in the eigenstates that 
causes spin decay through phonon emission, shown to be 
the dominant spin decoherence source for quantum dots. 
To minimize the effects of this mechanism the confining structure needs 
to have the minimum spin-orbit coupling.

In summary, we have shown that a guide structure in a 
semiconductor quantum well, defined by a spatially modulated Rashba 
spin-orbit coupling, gives rise to transverse-confined and longitudinal-propagating 
modes only for a given spin orientation. This feature characterizes the
proposed structures as spin guides that constitute natural paths to distinguish 
and drive the spin of the carriers within semiconductors. 
The feasibility of this spin-guiding mechanism is limited to low 
temperatures as the energies of the Rashba effect are in the range of 
a few meV's.
 
This work was supported by Grant No.\ BFM2002-03241 from DGI (Spain).

\begin{figure}[f]
\centerline{\psfig{figure=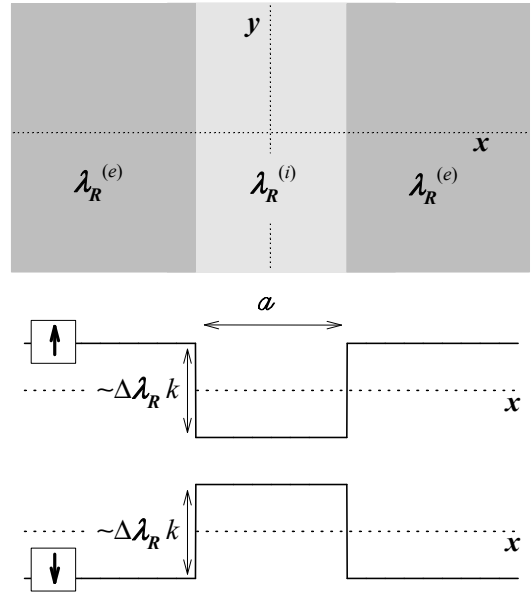,width=2.75in,clip=}}
\caption{Representation of the spatial modulation of 
spin-orbit coupling intensities in the heterostructure
containing the spin guide. 
The gray-scale plot shows an upper view of the $xy$ plane with 
the gray tones indicating different Rashba coupling intensities
(lighter gray corresponds to the spin guide).
The two lower plots display the effective potential in the 
simplified Hamiltonian, i.e., when $p_x\sigma_y$ is neglected, 
for $+x$ (up arrow) and $-x$ (down arrow) spin orientations.}
\end{figure}

\begin{figure}[f]
\centerline{\psfig{figure=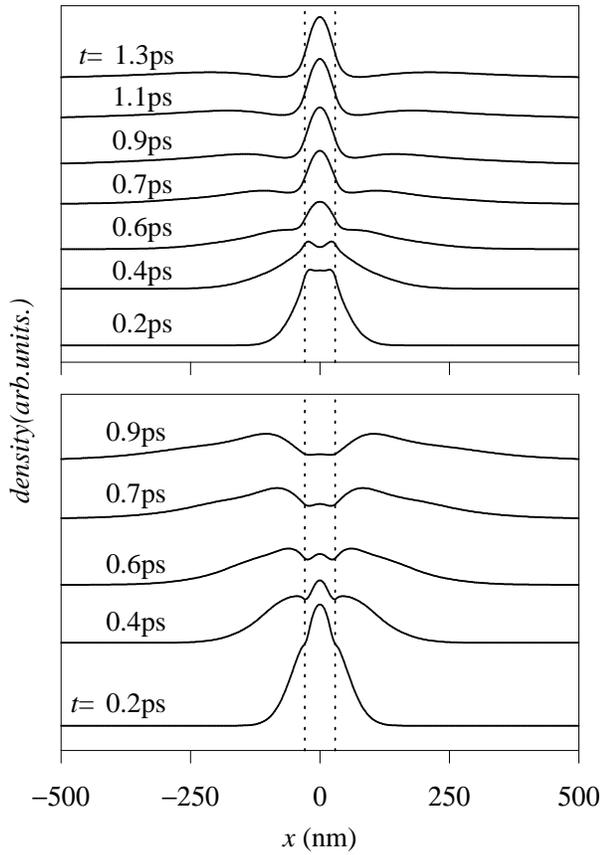,width=3.2in,clip=}}
\caption{Time evolution of the density for an initial Gaussian 
wavepacket with well-defined spin orientation for times up 
to $\approx 1$ ps. Upper panel corresponds to $+x$ initial spin 
direction while the lower one corresponds to $-x$.
The dotted lines indicate the position of the spin guide.}
\end{figure}

\begin{figure}[f]
\centerline{\psfig{figure=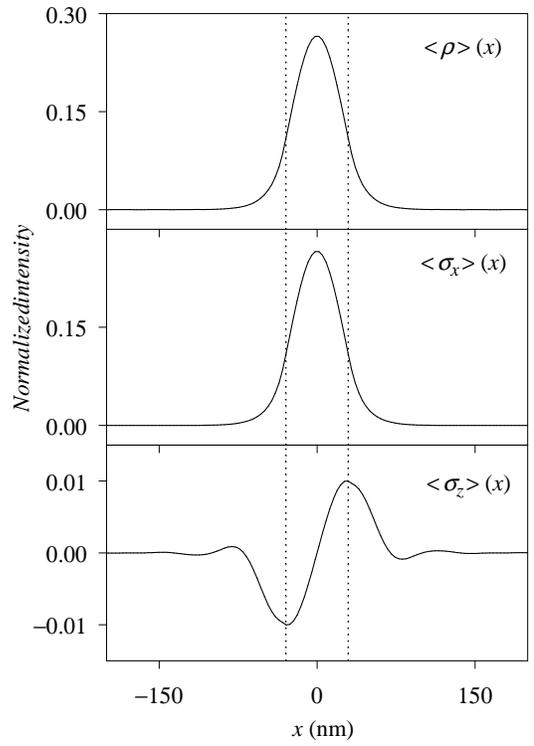,width=2.75in,clip=}}
\caption{Charge and spin densities corresponding to the
guide-confined mode. As in Fig.\ 2 the dotted lines mark 
the spin-guide edges.
Note the enlarged scale for $\langle\sigma_z\rangle(x)$.
}
\end{figure}

\begin{figure}[f]
\centerline{\psfig{figure=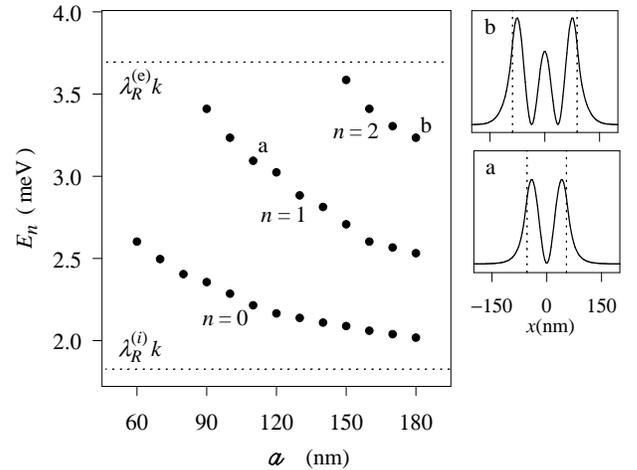,width=3.2in,clip=}}
\caption{ Confined mode energies as a function of the guide-width $a$.
Label $n$ indicates successive modes for a given $a$. The dotted lines
show the confinement edge energies. Right small plots display the 
probability densities $\langle \rho\rangle(x)$ for two 
selected modes labeled 'a' and 'b'.}
\end{figure}

\end{document}